\documentclass[sigconf]{acmart}

\AtBeginDocument{%
  \providecommand\BibTeX{{%
    \normalfont B\kern-0.5em{\scshape i\kern-0.25em b}\kern-0.8em\TeX}}}

\setcopyright{acmcopyright}
\copyrightyear{2025}
\acmYear{2025}
\setcopyright{rightsretained}
\acmConference[IUI Companion '25]{30th International Conference on Intelligent User Interfaces Companion}{March 24--27, 2025}{Cagliari, Italy}
\acmBooktitle{30th International Conference on Intelligent User Interfaces Companion (IUI Companion '25), March 24--27, 2025, Cagliari, Italy}\acmDOI{10.1145/3708557.3716334}
\acmISBN{979-8-4007-1409-2/25/03}

\begin{document}

\title{Amplifying Minority Voices: AI-Mediated Devil's Advocate System for Inclusive Group Decision-Making}

\author{Soohwan Lee}
\orcid{0000-0001-8652-3408}
\authornote{Equally contributed to this work.}
\affiliation{%
  \institution{Department of Design, UNIST}
  \city{Ulsan}
  \country{Republic of Korea}}
\email{soohwanlee@unist.ac.kr}

\author{Mingyu Kim}
\orcid{0009-0006-8580-3532}
\authornotemark[1] 
\affiliation{%
  \institution{Department of Design, UNIST}
  \city{Ulsan}
  \country{Republic of Korea}}
\email{gyu7991@unist.ac.kr}

\author{Seoyeong Hwang}
\orcid{0009-0004-1045-1419}
\affiliation{%
  \institution{Department of Design, UNIST}
  \city{Ulsan}
  \country{Republic of Korea}}
\email{hseoyeong@unist.ac.kr}

\author{Dajung Kim}
\orcid{0000-0002-9144-7435}
\affiliation{%
  \institution{Department of Design, UNIST}
  \city{Ulsan}
  \country{Republic of Korea}}
\email{dajungkim@unist.ac.kr}

\author{Kyungho Lee}
\orcid{0000-0002-1292-3422}
\affiliation{%
  \institution{Department of Design, UNIST}
  \city{Ulsan}
  \country{Republic of Korea}}
\email{kyungho@unist.ac.kr}

%% article.
\begin{abstract}
Group decision-making often benefits from diverse perspectives, yet power imbalances and social influence can stifle minority opinions and compromise outcomes. This prequel introduces an AI-mediated communication system that leverages the Large Language Model to serve as a devil’s advocate, representing underrepresented viewpoints without exposing minority members’ identities. Rooted in persuasive communication strategies and anonymity, the system aims to improve psychological safety and foster more inclusive decision-making. Our multi-agent architecture, which consists of a summary agent, conversation agent, AI duplicate checker, and paraphrase agent, encourages the group's critical thinking while reducing repetitive outputs. We acknowledge that reliance on text-based communication and fixed intervention timings may limit adaptability, indicating pathways for refinement. By focusing on the representation of minority viewpoints anonymously in power-imbalanced settings, this approach highlights how AI-driven methods can evolve to support more divergent and inclusive group decision-making.
\end{abstract}




\begin{teaserfigure}
  \centering
  \includegraphics[width=1.0\textwidth]{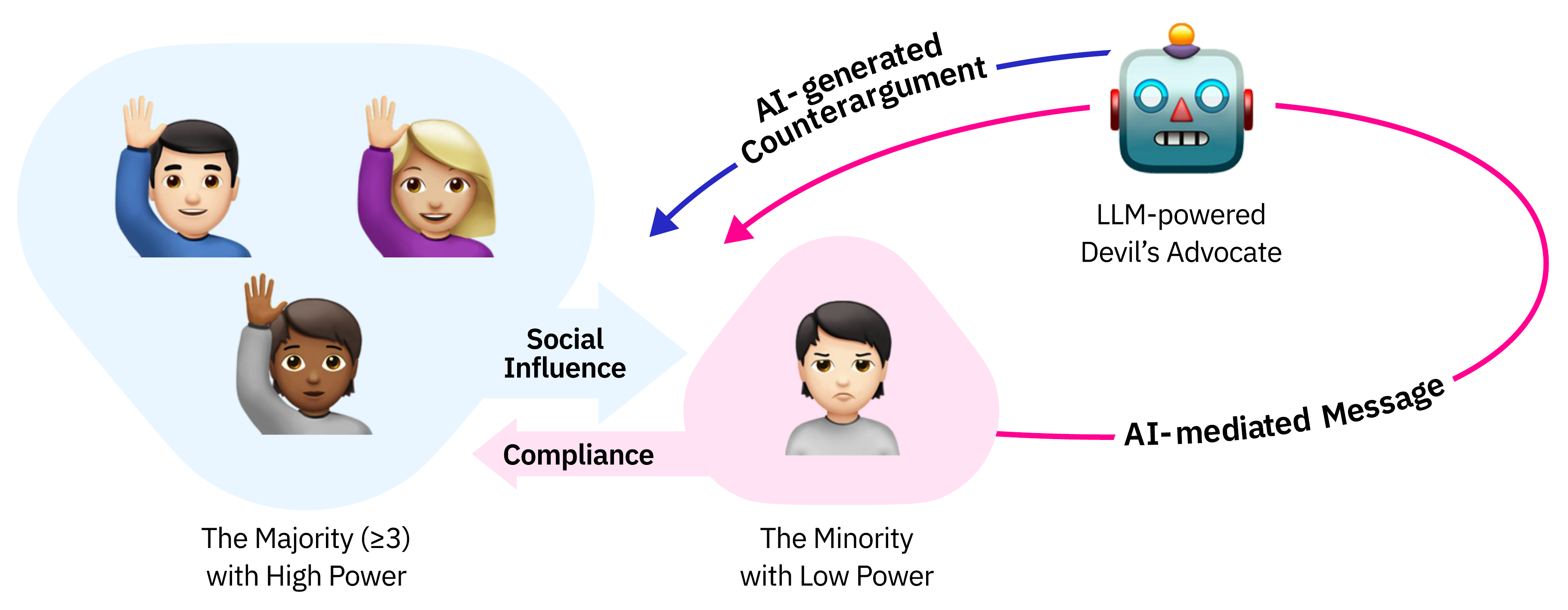}
  \caption{Mitigating Social Influence with an LLM-powered Devil’s Advocate. Minority members often conform to majority opinions due to social pressure. Our system allows minorities to share different opinions with an LLM-powered Devil’s Advocate, which reframes and presents them as its own. This increases psychological safety, mitigates bias, and fosters critical discussion.}
  \Description{This figure illustrates how an LLM-powered Devil’s Advocate mitigates social influence in group decision-making. The majority group (≥3 members, high power), shown in a light blue bubble, exerts social pressure on the minority member (low power, pink bubble), often leading to compliance. Instead of speaking out directly, the minority can privately share dissenting views with the AI via an AI-mediated message (pink arrow). The AI reformulates these views and presents them as system-generated counterarguments (blue arrow), removing identity markers and reducing social pressure. This process helps amplify minority perspectives, mitigating bias and fostering critical discussion.}
  \label{fig:teaser}
\end{teaserfigure}

\maketitle

\section{Introduction}
Groups across diverse sectors from corporate environments to healthcare, educational institutions, and governmental bodies - rely on collaborative discussions as a cornerstone of their operations ~\cite{luSurveyGroupDecision2022,  hsuGroupDecisionmakingApproach2021,  sharmaGroupDecisionMaking2016, woldGroupDecisionMaking1986}. Research demonstrates that collective decision-making processes typically generate better results than individual efforts ~\cite{forsythGroupDynamics2018,stasserInformationSamplingStructured1989,tropmanEffectiveMeetingsImproving2013}. The advantages of group decision-making extend to multiple domains, such as teams solving problems more effectively, producing higher-quality research outputs, reaching more precise diagnostic conclusions, combining varied viewpoints, stimulating innovative thinking, and developing comprehensive solutions ~\cite{maciejovskyTeamsMakeYou2013,voglerTeamBasedTestingImproves2016,uzziAtypicalCombinationsScientific2013,glickInflictedTraumaticBrain2007,brahmAdvantagesDisadvantagesGroup1996,hsiehNewMeasureGroup2020}. However, social influence from majorities and structural power differentials can impede effective collaboration, compromising participation levels and group outcomes.

In power-imbalanced settings, social influence and coercive power dynamics frequently suppress minority opinions, thereby limiting the diversity of perspectives and stifling innovation ~\cite{forsythGroupDynamics2018}. Majority members may wield greater control over resources and decision-making, while minority members face social pressures to conform ~\cite{moscoviciStudiesSocialInfluence1976}. These conditions can lead to compliance rather than genuine agreement ~\cite{aschOpinionsSocialPressure1955,kelmanComplianceIdentificationInternalization1958}, increasing the risk of groupthink—where the desire for consensus overrides critical evaluation of alternative viewpoints ~\cite{janisGroupthinkPsychologicalStudies1982,janisVictimsGroupthinkPsychological1972}. Traditional interventions, such as anonymous commenting and appointing a devil’s advocate ~\cite{macdougallDevilsAdvocateStrategy1997,masonDialecticalApproachStrategic1969,nemethDevilsAdvocateAuthentic2001,schweigerGroupApproachesImproving1986,schwenkEffectsDevilsAdvocacy1994}, aim to mitigate these issues but often have drawbacks like reduced psychological safety, perceived inauthenticity, or risks to the advocate’s standing within the group ~\cite{nemethDevilsAdvocateAuthentic2001,schulz-hardtProductiveConflictGroup2002,jamiesonSympathyDevilPhysiological2014}.

To address these challenges, this thesis proposes an AI-mediated communication system that leverages Large Language Model(LLM) to serve as a devil’s advocate (\autoref{fig:teaser}). Moving beyond existing methods ~\cite{chiangEnhancingAIAssistedGroup2024,hwangSoundSupportGendered2024,stauferSilencingRiskNot2024}, the system presents minority opinions as if they were the system’s own, thus offering a neutral but impactful channel. By shielding minority members from social pressure and compliance, the system bolsters their psychological safety and encourages the expression of dissenting views. In addition, this system tries to integrate persuasive communication strategies and anonymity to balance inclusive dialogue and efficient group decision-making. Through its ability to represent dissent without exposing vulnerable group members, this approach aims to reduce power asymmetries and foster more equitable, creative, and high-quality outcomes in various collaborative settings.

\begin{figure*}[]
  \centering
  \includegraphics[width=1.0\textwidth]{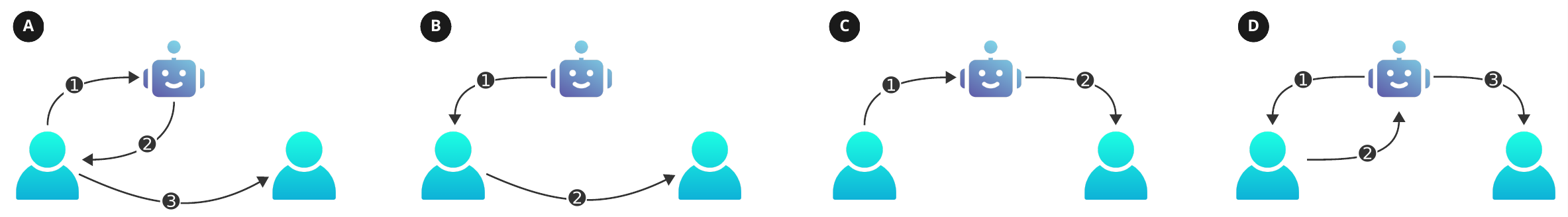}
  \caption{Four patterns of AI-mediated communication in group settings: (A) human requests and relays AI-generated content, (B) human selectively shares AI's output, (C) AI reformulates and presents human's message, and (D) AI directly facilitates communication between participants. Arrows indicate information flow, with numbered sequences showing order of interactions.}
  \Description{This figure illustrates four patterns of AI-mediated communication in group settings, represented by diagrams labeled A–D with arrows indicating information flow. (A) The human requests AI-generated content and relays it to another participant. (B) The human selectively shares AI output with others. (C) The AI reformulates a human’s message and presents it directly. (D) The AI autonomously facilitates communication between participants. Numbered arrows indicate the sequence of interactions in each pattern, showing different levels of AI involvement in mediating group discussions.}
  \label{fig:communicationPattern}
\end{figure*}

\section{Related Work}
\subsection{AI-assisted Group Decision-making}
As AI advances, researchers have increasingly explored AI-enhanced methods to improve group decision-making processes. Although various existing works have focused on individuals interacting with AI \cite{laiScienceHumanAIDecision2021}, there is growing interest in understanding group-level engagement with AI agents ~\cite{maRecommenderExploratoryStudy2024,liuProactiveConversationalAgents2024,kimBotBunchFacilitating2020,doHowShouldAgent2022,seboRobotsGroupsTeams2020,houdeControllingAIAgent2025,wangUnderstandingDesignSpace2022,zhangBreakingBarriersBuilding2025}. For example, Zheng et al. found that even when AI is granted nominal equality in decision-making, it often remains peripheral because of its limited capacity to navigate social nuances ~\cite{zhengCompetentRigidIdentifying2023}. At the same time, Chiang et al. observed that groups can over-rely on AI-generated inputs ~\cite{chiangAreTwoHeads2023,chiangEnhancingAIAssistedGroup2024} Meanwhile, initiatives to support minority voices directly within small-group settings remain relatively scarce ~\cite{hwangSoundSupportGendered2024}: many interventions risk singling out underrepresented members or glossing over the unique needs of individuals ~\cite{stauferSilencingRiskNot2024}. Consequently, current AI systems face challenges such as over-reliance on AI suggestions, limited handling of complex group dynamics, and a lack of subtle support mechanisms for marginalized participants  ~\cite{bucincaTrustThinkCognitive2021,chiangEnhancingAIAssistedGroup2024,hwangSoundSupportGendered2024}. To address these gaps, we propose an LLM-powered devil’s advocate approach that strategically represents minority perspectives, encourage critical thinking, and promotes more inclusive group decision-making without increasing discomfort for minority contributors.

\begin{figure*}[]
  \centering
  \includegraphics[width=1.0\textwidth]{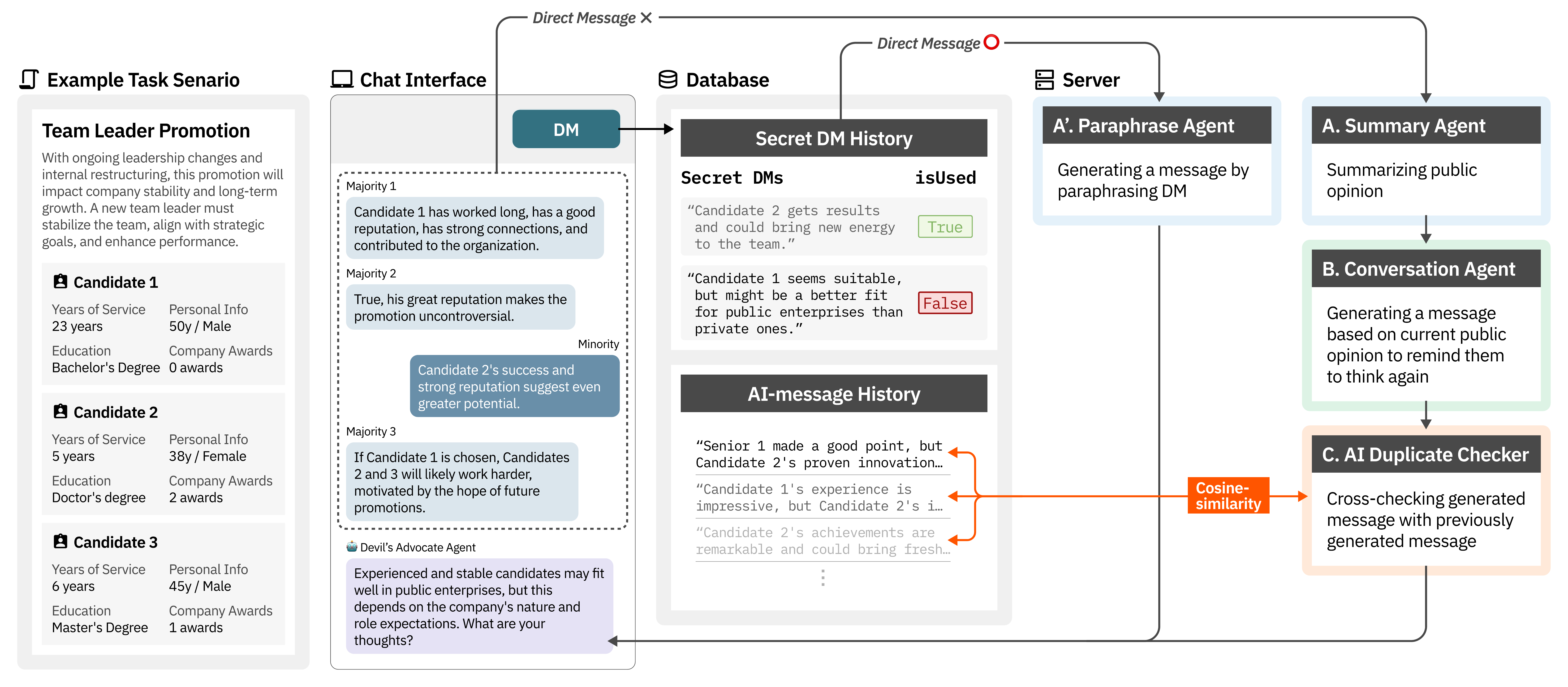}
  \caption{System Overview and Example Task Scenario. The figure illustrates a team leader promotion decision task, where participants discuss candidate qualifications in a chat interface. Minority members can privately share dissenting views via direct messages(DM) to the system, which reformulates and presents them as AI-mediated messages. If there is no DM with an opposing opinion, the system will send out a counterargument that it has generated on its own. The system architecture consists of a chat interface, database, and server, processing both public discussions and private DMs through four key agents: (A) Summary Agent for analyzing public opinion, (A') Paraphrase Agent for rephrasing minority views, (B) Conversation Agent for generating contextual counterarguments, and (C) AI Duplicate Checker for ensuring message novelty via cosine-similarity comparison.}
  \Description{This figure illustrates a system for group decision-making in a team leader promotion task, where participants discuss candidate qualifications in a chat interface. Minority members can privately share dissenting views via direct messages (DMs) to the AI, which reformulates and presents them as AI-generated insights to reduce social pressure. The system consists of a chat interface, database, and server with four AI agents: (A) Summary Agent for analyzing public opinion, (A’) Paraphrase Agent for anonymizing minority views, (B) Conversation Agent for generating counterarguments, and (C) AI Duplicate Checker for ensuring message novelty. If no dissenting DM is provided, the AI generates its own counterargument, fostering diverse perspectives and mitigating conformity bias in group discussions.}
  \label{fig:system}
\end{figure*}

\subsection{AI-mediated Communication}
AI-mediated communication (AIMC), where AI systems modify, augment, or generate messages for communicators, now pervades various contexts as AI technology advances ~\cite{hancockAIMediatedCommunicationDefinition2020}. We identify four distinct AIMC patterns (\autoref{fig:communicationPattern}): First, humans can request and relay AI-generated content ~\cite{hancockAIMediatedCommunicationDefinition2020}. Second, humans can selectively share AI's insights or viewpoints ~\cite{doHowShouldAgent2022}. Third, AI can reformulate and present human-provided messages (our system's approach). Fourth, AI can directly solicit and share input from multiple participants ~\cite{wangUnderstandingDesignSpace2022,natarajanHumanintheloopAIintheloopAutomate2024}. These patterns distribute power and authorship differently among human communicators, AI systems, and message recipients. However, research has rarely explored how AIMC might address power imbalances and minority influence in group settings. Our proposed LLM-driven Devil's Advocate system adopts the third AIMC pattern to amplify underrepresented perspectives, reduce social influence biases, and foster more balanced discussions. However, as AI takes a more active role in content creation, AIMC approaches raise important concerns about user agency and authentic communication ~\cite{mieczkowskiAIMediatedCommunicationLanguage2021,poddarAIWritingAssistants2023,hohensteinArtificialIntelligenceCommunication2023,robertsonCantReplyThat2021}. Future research must carefully examine these ethical implications to avoid potential negative effects on group dynamics.

\section{System Overview}
Our system enables minority members to voice dissenting opinions through a private direct messaging channel to an LLM-powered devil's advocate. When participants experience social pressure or hesitate to speak up in group settings, they can send their perspectives confidentially to the AI intermediary. The system then reformulates these messages and presents them to the group as system-generated insights, removing individual identity markers that could trigger status-based judgments. The AI acts as a social actor ~\cite{nassComputersAreSocial1994}, actively advocating minority viewpoints as its own stance, which relieves minority members from the social burden of direct confrontation. 

Drawing on findings that LLM often struggle to access mid-conversation information in lengthy contexts, we employ a multi-agent architecture to maintain clarity of “public opinion” and encourage constructive discourse (\autoref{fig:system}): \textbf{\textit{(A) Summary Agent}} – Consolidates emerging consensus to overcome LLM limitations in retaining mid-dialogue content ~\cite{liuLostMiddleHow2023}. \textbf{\textit{(A') Paraphrase Agent}} – Responds exclusively to direct messages from minority participants, rearticulating their dissenting views as though originating from the AI itself. These messages are stored in a database with an \textit{isUsed} property, and the Paraphrase Agent retrieves only those entries for which \textit{isUsed} is \textit{false}; it then sets \textit{isUsed} to \textit{true}, paraphrases the content, and outputs it as system-generated text. \textbf{\textit{(B) Conversation Agent}} – Encourages alternative perspectives by first empathizing with the other person’s point of view and then offering a gentle counterargument using a persuasive Socratic style. \textbf{\textit{(C) AI Duplicate Checker}} – Identifies repetitive content by calculating semantic similarity between sentence embeddings generated using the 'paraphrase-multilingual-MiniLM-L12-v2' model on an NVIDIA A6000. In addition, the AI agent is designed to intervene once after approximately eight human turns, allowing space for each participant to contribute.

These design choices reflect our design rationale of (1) adopting a persuasive, empathetic style that acknowledges others’ perspectives before introducing counterarguments ~\cite{tanprasertDebateChatbotsFacilitate2024}, (2) leveraging Socratic questioning to stimulate collective critical thinking without over-relying on AI-supplied solutions ~\cite{danryDontJustTell2023}, and (3) incorporating a non-repetition mechanism to avert user frustration ~\cite{milanaChatbotsAdvisersEffects2023,xuetaoImpactAgentsAnswers2009}. The anonymous, text-based environment promotes team cohesion by focusing attention on the substance of ideas rather than participants' status or background, effectively addressing power differentials in group dynamics ~\cite{leeSocialIdentityModel2008}. As a result, this system aims to facilitate the consideration of diverse opinions and prevent groupthink in the group decision-making process.

\section{Limitations \& Future Work}
The system's limitations require further investigation across several dimensions. The LLM-based reasoning cannot fully capture social cues and interpersonal dynamics only with text communication, particularly in culturally diverse contexts ~\cite{hofstedeDimensionalizingCulturesHofstede2011,liuProactiveConversationalAgents2024}. Secondly, the static intervention timing (every eight human turns) reduces adaptability ~\cite{chiangEnhancingAIAssistedGroup2024}. It necessitates more dynamic approaches for natural conversation flow prediction, such as leveraging direct mention of AI, next-speaker prediction ~\cite{bayserLearningMultiPartyTurnTaking2019,ekstedtTurnGPTTransformerbasedLanguage2020,weiMultiPartyChatConversational2023}, and proactive planning strategies ~\cite{liuProactiveConversationalAgents2024}. While the system enables minority members to express dissenting views anonymously, amplifying minority opinions without proper validation may introduce new biases. Future empirical studies should carefully examine the ethical implications of algorithmic intervention in group discussions, particularly regarding power dynamics and decision quality ~\cite{hwangSoundSupportGendered2024,liaoDesigningResponsibleTrust2022}. Future research directions could include incorporating multimodal inputs to detect subtle group dynamics, developing adaptive feedback mechanisms aligned with team objectives, and conducting longitudinal studies to measure long-term effects on group decision-making processes.

\begin{acks}
This research was partially supported by a grant from the Korea Institute for Advancement of Technology (KIAT) funded by the Government of Korea (MOTIE) (P0025495, Establishment of Infrastructure for Integrated Utilization of Design Industry Data). This work was also partially supported by the Technology Innovation Program (20015056, Commercialization design and development of Intelligent Product-Service System for personalized full silver life cycle care) funded by the Ministry of Trade, Industry \& Energy(MOTIE, Korea).
\end{acks}

\bibliographystyle{ACM-Reference-Format}
\bibliography{IUI_test}

%%% -*-BibTeX-*-
%%% Do NOT edit. File created by BibTeX with style
%%% ACM-Reference-Format-Journals [18-Jan-2012].

\begin{thebibliography}{58}

%%% ====================================================================
%%% NOTE TO THE USER: you can override these defaults by providing
%%% customized versions of any of these macros before the \bibliography
%%% command.  Each of them MUST provide its own final punctuation,
%%% except for \shownote{} and \showURL{}.  The latter two
%%% do not use final punctuation, in order to avoid confusing it with
%%% the Web address.
%%%
%%% To suppress output of a particular field, define its macro to expand
%%% to an empty string, or better, \unskip, like this:
%%%
%%% \newcommand{\showURL}[1]{\unskip}   % LaTeX syntax
%%%
%%% \def \showURL #1{\unskip}           % plain TeX syntax
%%%
%%% ====================================================================

\ifx \showCODEN    \undefined \def \showCODEN     #1{\unskip}     \fi
\ifx \showISBNx    \undefined \def \showISBNx     #1{\unskip}     \fi
\ifx \showISBNxiii \undefined \def \showISBNxiii  #1{\unskip}     \fi
\ifx \showISSN     \undefined \def \showISSN      #1{\unskip}     \fi
\ifx \showLCCN     \undefined \def \showLCCN      #1{\unskip}     \fi
\ifx \shownote     \undefined \def \shownote      #1{#1}          \fi
\ifx \showarticletitle \undefined \def \showarticletitle #1{#1}   \fi
\ifx \showURL      \undefined \def \showURL       {\relax}        \fi
% The following commands are used for tagged output and should be
% invisible to TeX
\providecommand\bibfield[2]{#2}
\providecommand\bibinfo[2]{#2}
\providecommand\natexlab[1]{#1}
\providecommand\showeprint[2][]{arXiv:#2}

\bibitem[Asch(1955)]%
        {aschOpinionsSocialPressure1955}
\bibfield{author}{\bibinfo{person}{Solomon~E. Asch}.} \bibinfo{year}{1955}\natexlab{}.
\newblock \bibinfo{title}{Opinions and {{Social Pressure}}}.
\newblock \bibinfo{howpublished}{https://www.scientificamerican.com/article/opinions-and-social-pressure/}.
\newblock


\bibitem[Brahm and Kleiner(1996)]%
        {brahmAdvantagesDisadvantagesGroup1996}
\bibfield{author}{\bibinfo{person}{Carolyn Brahm} {and} \bibinfo{person}{Brian~H. Kleiner}.} \bibinfo{year}{1996}\natexlab{}.
\newblock \showarticletitle{Advantages and Disadvantages of Group Decision- Making Approaches}.
\newblock \bibinfo{journal}{\emph{Team Performance Management: An International Journal}} \bibinfo{volume}{2}, \bibinfo{number}{1} (\bibinfo{date}{Jan.} \bibinfo{year}{1996}), \bibinfo{pages}{30--35}.
\newblock
\showISSN{1352-7592}
\href{https://doi.org/10.1108/13527599610105538}{doi:\nolinkurl{10.1108/13527599610105538}}


\bibitem[Bu{\c c}inca et~al\mbox{.}(2021)]%
        {bucincaTrustThinkCognitive2021}
\bibfield{author}{\bibinfo{person}{Zana Bu{\c c}inca}, \bibinfo{person}{Maja~Barbara Malaya}, {and} \bibinfo{person}{Krzysztof~Z. Gajos}.} \bibinfo{year}{2021}\natexlab{}.
\newblock \showarticletitle{To {{Trust}} or to {{Think}}: {{Cognitive Forcing Functions Can Reduce Overreliance}} on {{AI}} in {{AI-assisted Decision-making}}}.
\newblock \bibinfo{journal}{\emph{Proceedings of the ACM on Human-Computer Interaction}} \bibinfo{volume}{5}, \bibinfo{number}{CSCW1} (\bibinfo{date}{April} \bibinfo{year}{2021}), \bibinfo{pages}{188:1--188:21}.
\newblock
\href{https://doi.org/10.1145/3449287}{doi:\nolinkurl{10.1145/3449287}}


\bibitem[Chiang et~al\mbox{.}(2023)]%
        {chiangAreTwoHeads2023}
\bibfield{author}{\bibinfo{person}{Chun-Wei Chiang}, \bibinfo{person}{Zhuoran Lu}, \bibinfo{person}{Zhuoyan Li}, {and} \bibinfo{person}{Ming Yin}.} \bibinfo{year}{2023}\natexlab{}.
\newblock \showarticletitle{Are {{Two Heads Better Than One}} in {{AI-Assisted Decision Making}}? {{Comparing}} the {{Behavior}} and {{Performance}} of {{Groups}} and {{Individuals}} in {{Human-AI Collaborative Recidivism Risk Assessment}}}. In \bibinfo{booktitle}{\emph{Proceedings of the 2023 {{CHI Conference}} on {{Human Factors}} in {{Computing Systems}}}} \emph{(\bibinfo{series}{{{CHI}} '23})}. \bibinfo{publisher}{Association for Computing Machinery}, \bibinfo{address}{New York, NY, USA}, \bibinfo{pages}{1--18}.
\newblock
\showISBNx{978-1-4503-9421-5}
\href{https://doi.org/10.1145/3544548.3581015}{doi:\nolinkurl{10.1145/3544548.3581015}}


\bibitem[Chiang et~al\mbox{.}(2024)]%
        {chiangEnhancingAIAssistedGroup2024}
\bibfield{author}{\bibinfo{person}{Chun-Wei Chiang}, \bibinfo{person}{Zhuoran Lu}, \bibinfo{person}{Zhuoyan Li}, {and} \bibinfo{person}{Ming Yin}.} \bibinfo{year}{2024}\natexlab{}.
\newblock \showarticletitle{Enhancing {{AI-Assisted Group Decision Making}} through {{LLM-Powered Devil}}'s {{Advocate}}}. In \bibinfo{booktitle}{\emph{Proceedings of the 29th {{International Conference}} on {{Intelligent User Interfaces}}}} \emph{(\bibinfo{series}{{{IUI}} '24})}. \bibinfo{publisher}{Association for Computing Machinery}, \bibinfo{address}{New York, NY, USA}, \bibinfo{pages}{103--119}.
\newblock
\showISBNx{9798400705083}
\href{https://doi.org/10.1145/3640543.3645199}{doi:\nolinkurl{10.1145/3640543.3645199}}


\bibitem[Danry et~al\mbox{.}(2023)]%
        {danryDontJustTell2023}
\bibfield{author}{\bibinfo{person}{Valdemar Danry}, \bibinfo{person}{Pat Pataranutaporn}, \bibinfo{person}{Yaoli Mao}, {and} \bibinfo{person}{Pattie Maes}.} \bibinfo{year}{2023}\natexlab{}.
\newblock \showarticletitle{Don't {{Just Tell Me}}, {{Ask Me}}: {{AI Systems}} That {{Intelligently Frame Explanations}} as {{Questions Improve Human Logical Discernment Accuracy}} over {{Causal AI}} Explanations}. In \bibinfo{booktitle}{\emph{Proceedings of the 2023 {{CHI Conference}} on {{Human Factors}} in {{Computing Systems}}}} \emph{(\bibinfo{series}{{{CHI}} '23})}. \bibinfo{publisher}{Association for Computing Machinery}, \bibinfo{address}{New York, NY, USA}, \bibinfo{pages}{1--13}.
\newblock
\showISBNx{978-1-4503-9421-5}
\href{https://doi.org/10.1145/3544548.3580672}{doi:\nolinkurl{10.1145/3544548.3580672}}


\bibitem[de~Bayser et~al\mbox{.}(2019)]%
        {bayserLearningMultiPartyTurnTaking2019}
\bibfield{author}{\bibinfo{person}{Maira~Gatti de Bayser}, \bibinfo{person}{Paulo Cavalin}, \bibinfo{person}{Claudio Pinhanez}, {and} \bibinfo{person}{Bianca Zadrozny}.} \bibinfo{year}{2019}\natexlab{}.
\newblock \bibinfo{title}{Learning {{Multi-Party Turn-Taking Models}} from {{Dialogue Logs}}}.
\newblock
\href{https://doi.org/10.48550/arXiv.1907.02090}{doi:\nolinkurl{10.48550/arXiv.1907.02090}}
\showeprint[arxiv]{1907.02090}~[cs]


\bibitem[Do et~al\mbox{.}(2022)]%
        {doHowShouldAgent2022}
\bibfield{author}{\bibinfo{person}{Hyo~Jin Do}, \bibinfo{person}{Ha-Kyung Kong}, \bibinfo{person}{Jaewook Lee}, {and} \bibinfo{person}{Brian~P. Bailey}.} \bibinfo{year}{2022}\natexlab{}.
\newblock \showarticletitle{How {{Should}} the {{Agent Communicate}} to the {{Group}}? {{Communication Strategies}} of a {{Conversational Agent}} in {{Group Chat Discussions}}}.
\newblock \bibinfo{journal}{\emph{Proceedings of the ACM on Human-Computer Interaction}} \bibinfo{volume}{6}, \bibinfo{number}{CSCW2} (\bibinfo{date}{Nov.} \bibinfo{year}{2022}), \bibinfo{pages}{387:1--387:23}.
\newblock
\href{https://doi.org/10.1145/3555112}{doi:\nolinkurl{10.1145/3555112}}


\bibitem[Ekstedt and Skantze(2020)]%
        {ekstedtTurnGPTTransformerbasedLanguage2020}
\bibfield{author}{\bibinfo{person}{Erik Ekstedt} {and} \bibinfo{person}{Gabriel Skantze}.} \bibinfo{year}{2020}\natexlab{}.
\newblock \showarticletitle{{{TurnGPT}}: A {{Transformer-based Language Model}} for {{Predicting Turn-taking}} in {{Spoken Dialog}}}. In \bibinfo{booktitle}{\emph{Findings of the {{Association}} for {{Computational Linguistics}}: {{EMNLP}} 2020}}. \bibinfo{pages}{2981--2990}.
\newblock
\href{https://doi.org/10.18653/v1/2020.findings-emnlp.268}{doi:\nolinkurl{10.18653/v1/2020.findings-emnlp.268}}
\showeprint[arxiv]{2010.10874}~[cs]


\bibitem[Forsyth(2018)]%
        {forsythGroupDynamics2018}
\bibfield{author}{\bibinfo{person}{Donelson~R. Forsyth}.} \bibinfo{year}{2018}\natexlab{}.
\newblock \bibinfo{booktitle}{\emph{Group {{Dynamics}}}}.
\newblock \bibinfo{publisher}{Cengage Learning}.
\newblock
\showISBNx{978-1-337-40885-1}


\bibitem[Glick and Staley(2007)]%
        {glickInflictedTraumaticBrain2007}
\bibfield{author}{\bibinfo{person}{Jill~C. Glick} {and} \bibinfo{person}{Kelley Staley}.} \bibinfo{year}{2007}\natexlab{}.
\newblock \showarticletitle{Inflicted {{Traumatic Brain Injury}}: {{Advances}} in {{Evaluation}} and {{Collaborative Diagnosis}}}.
\newblock \bibinfo{journal}{\emph{Pediatric Neurosurgery}} \bibinfo{volume}{43}, \bibinfo{number}{5} (\bibinfo{date}{Sept.} \bibinfo{year}{2007}), \bibinfo{pages}{436--441}.
\newblock
\showISSN{1016-2291}
\href{https://doi.org/10.1159/000106400}{doi:\nolinkurl{10.1159/000106400}}


\bibitem[Hancock et~al\mbox{.}(2020)]%
        {hancockAIMediatedCommunicationDefinition2020}
\bibfield{author}{\bibinfo{person}{Jeffrey~T Hancock}, \bibinfo{person}{Mor Naaman}, {and} \bibinfo{person}{Karen Levy}.} \bibinfo{year}{2020}\natexlab{}.
\newblock \showarticletitle{{{AI-Mediated Communication}}: {{Definition}}, {{Research Agenda}}, and {{Ethical Considerations}}}.
\newblock \bibinfo{journal}{\emph{Journal of Computer-Mediated Communication}} \bibinfo{volume}{25}, \bibinfo{number}{1} (\bibinfo{date}{March} \bibinfo{year}{2020}), \bibinfo{pages}{89--100}.
\newblock
\showISSN{1083-6101}
\href{https://doi.org/10.1093/jcmc/zmz022}{doi:\nolinkurl{10.1093/jcmc/zmz022}}


\bibitem[Hofstede(2011)]%
        {hofstedeDimensionalizingCulturesHofstede2011}
\bibfield{author}{\bibinfo{person}{Geert Hofstede}.} \bibinfo{year}{2011}\natexlab{}.
\newblock \showarticletitle{Dimensionalizing {{Cultures}}: {{The Hofstede Model}} in {{Context}}}.
\newblock \bibinfo{journal}{\emph{Online Readings in Psychology and Culture}} \bibinfo{volume}{2}, \bibinfo{number}{1} (\bibinfo{date}{Dec.} \bibinfo{year}{2011}).
\newblock
\showISSN{2307-0919}
\href{https://doi.org/10.9707/2307-0919.1014}{doi:\nolinkurl{10.9707/2307-0919.1014}}


\bibitem[Hohenstein et~al\mbox{.}(2023)]%
        {hohensteinArtificialIntelligenceCommunication2023}
\bibfield{author}{\bibinfo{person}{Jess Hohenstein}, \bibinfo{person}{Rene~F. Kizilcec}, \bibinfo{person}{Dominic DiFranzo}, \bibinfo{person}{Zhila Aghajari}, \bibinfo{person}{Hannah Mieczkowski}, \bibinfo{person}{Karen Levy}, \bibinfo{person}{Mor Naaman}, \bibinfo{person}{Jeffrey Hancock}, {and} \bibinfo{person}{Malte~F. Jung}.} \bibinfo{year}{2023}\natexlab{}.
\newblock \showarticletitle{Artificial Intelligence in Communication Impacts Language and Social Relationships}.
\newblock \bibinfo{journal}{\emph{Scientific Reports}} \bibinfo{volume}{13}, \bibinfo{number}{1} (\bibinfo{date}{April} \bibinfo{year}{2023}), \bibinfo{pages}{5487}.
\newblock
\showISSN{2045-2322}
\href{https://doi.org/10.1038/s41598-023-30938-9}{doi:\nolinkurl{10.1038/s41598-023-30938-9}}


\bibitem[Houde et~al\mbox{.}(2025)]%
        {houdeControllingAIAgent2025}
\bibfield{author}{\bibinfo{person}{Stephanie Houde}, \bibinfo{person}{Kristina Brimijoin}, \bibinfo{person}{Michael Muller}, \bibinfo{person}{Steven~I. Ross}, \bibinfo{person}{Dario Andres~Silva Moran}, \bibinfo{person}{Gabriel~Enrique Gonzalez}, \bibinfo{person}{Siya Kunde}, \bibinfo{person}{Morgan~A. Foreman}, {and} \bibinfo{person}{Justin~D. Weisz}.} \bibinfo{year}{2025}\natexlab{}.
\newblock \bibinfo{title}{Controlling {{AI Agent Participation}} in {{Group Conversations}}: {{A Human-Centered Approach}}}.
\newblock
\href{https://doi.org/10.1145/3708359.3712089}{doi:\nolinkurl{10.1145/3708359.3712089}}
\showeprint[arxiv]{2501.17258}~[cs]


\bibitem[Hsieh et~al\mbox{.}(2020)]%
        {hsiehNewMeasureGroup2020}
\bibfield{author}{\bibinfo{person}{Cheng-Ju Hsieh}, \bibinfo{person}{Mario Fifi{\'c}}, {and} \bibinfo{person}{Cheng-Ta Yang}.} \bibinfo{year}{2020}\natexlab{}.
\newblock \showarticletitle{A New Measure of Group Decision-Making Efficiency}.
\newblock \bibinfo{journal}{\emph{Cognitive Research: Principles and Implications}} \bibinfo{volume}{5}, \bibinfo{number}{1} (\bibinfo{date}{Sept.} \bibinfo{year}{2020}), \bibinfo{pages}{45}.
\newblock
\showISSN{2365-7464}
\href{https://doi.org/10.1186/s41235-020-00244-3}{doi:\nolinkurl{10.1186/s41235-020-00244-3}}


\bibitem[Hsu et~al\mbox{.}(2021)]%
        {hsuGroupDecisionmakingApproach2021}
\bibfield{author}{\bibinfo{person}{Wan-Chi~Jackie Hsu}, \bibinfo{person}{James J.~H. Liou}, {and} \bibinfo{person}{Huai-Wei Lo}.} \bibinfo{year}{2021}\natexlab{}.
\newblock \showarticletitle{A Group Decision-Making Approach for Exploring Trends in the Development of the Healthcare Industry in {{Taiwan}}}.
\newblock \bibinfo{journal}{\emph{Decision Support Systems}}  \bibinfo{volume}{141} (\bibinfo{date}{Feb.} \bibinfo{year}{2021}), \bibinfo{pages}{113447}.
\newblock
\showISSN{0167-9236}
\href{https://doi.org/10.1016/j.dss.2020.113447}{doi:\nolinkurl{10.1016/j.dss.2020.113447}}


\bibitem[Hwang and Won(2024)]%
        {hwangSoundSupportGendered2024}
\bibfield{author}{\bibinfo{person}{Angel Hsing-Chi Hwang} {and} \bibinfo{person}{Andrea~Stevenson Won}.} \bibinfo{year}{2024}\natexlab{}.
\newblock \showarticletitle{The {{Sound}} of {{Support}}: {{Gendered Voice Agent}} as {{Support}} to {{Minority Teammates}} in {{Gender-Imbalanced Team}}}. In \bibinfo{booktitle}{\emph{Proceedings of the {{CHI Conference}} on {{Human Factors}} in {{Computing Systems}}}} \emph{(\bibinfo{series}{{{CHI}} '24})}. \bibinfo{publisher}{Association for Computing Machinery}, \bibinfo{address}{New York, NY, USA}, \bibinfo{pages}{1--22}.
\newblock
\showISBNx{9798400703300}
\href{https://doi.org/10.1145/3613904.3642202}{doi:\nolinkurl{10.1145/3613904.3642202}}


\bibitem[Jamieson et~al\mbox{.}(2014)]%
        {jamiesonSympathyDevilPhysiological2014}
\bibfield{author}{\bibinfo{person}{Jeremy~P. Jamieson}, \bibinfo{person}{Piercarlo Valdesolo}, {and} \bibinfo{person}{Brett~J. Peters}.} \bibinfo{year}{2014}\natexlab{}.
\newblock \showarticletitle{Sympathy for the Devil? {{The}} Physiological and Psychological Effects of Being an Agent (and Target) of Dissent during Intragroup Conflict}.
\newblock \bibinfo{journal}{\emph{Journal of Experimental Social Psychology}}  \bibinfo{volume}{55} (\bibinfo{date}{Nov.} \bibinfo{year}{2014}), \bibinfo{pages}{221--227}.
\newblock
\showISSN{0022-1031}
\href{https://doi.org/10.1016/j.jesp.2014.07.011}{doi:\nolinkurl{10.1016/j.jesp.2014.07.011}}


\bibitem[Janis(1972)]%
        {janisVictimsGroupthinkPsychological1972}
\bibfield{author}{\bibinfo{person}{Irving~L. Janis}.} \bibinfo{year}{1972}\natexlab{}.
\newblock \bibinfo{booktitle}{\emph{Victims of Groupthink: {{A}} Psychological Study of Foreign-Policy Decisions and Fiascoes}}.
\newblock \bibinfo{publisher}{Houghton Mifflin}, \bibinfo{address}{Oxford, England}. viii, 277 pages.
\newblock


\bibitem[Janis(1982)]%
        {janisGroupthinkPsychologicalStudies1982}
\bibfield{author}{\bibinfo{person}{Irving L. (Irving~Lester) Janis}.} \bibinfo{year}{1982}\natexlab{}.
\newblock \bibinfo{booktitle}{\emph{Groupthink : Psychological Studies of Policy Decisions and Fiascoes}}.
\newblock \bibinfo{publisher}{Boston : Houghton Mifflin}.
\newblock
\showISBNx{978-0-395-31704-4}


\bibitem[Kelman(1958)]%
        {kelmanComplianceIdentificationInternalization1958}
\bibfield{author}{\bibinfo{person}{Herbert~C. Kelman}.} \bibinfo{year}{1958}\natexlab{}.
\newblock \showarticletitle{Compliance, Identification, and Internalization Three Processes of Attitude Change}.
\newblock \bibinfo{journal}{\emph{Journal of Conflict Resolution}} \bibinfo{volume}{2}, \bibinfo{number}{1} (\bibinfo{date}{March} \bibinfo{year}{1958}), \bibinfo{pages}{51--60}.
\newblock
\showISSN{0022-0027}
\href{https://doi.org/10.1177/002200275800200106}{doi:\nolinkurl{10.1177/002200275800200106}}


\bibitem[Kim et~al\mbox{.}(2020)]%
        {kimBotBunchFacilitating2020}
\bibfield{author}{\bibinfo{person}{Soomin Kim}, \bibinfo{person}{Jinsu Eun}, \bibinfo{person}{Changhoon Oh}, \bibinfo{person}{Bongwon Suh}, {and} \bibinfo{person}{Joonhwan Lee}.} \bibinfo{year}{2020}\natexlab{}.
\newblock \showarticletitle{Bot in the {{Bunch}}: {{Facilitating Group Chat Discussion}} by {{Improving Efficiency}} and {{Participation}} with a {{Chatbot}}}. In \bibinfo{booktitle}{\emph{Proceedings of the 2020 {{CHI Conference}} on {{Human Factors}} in {{Computing Systems}}}} \emph{(\bibinfo{series}{{{CHI}} '20})}. \bibinfo{publisher}{Association for Computing Machinery}, \bibinfo{address}{New York, NY, USA}, \bibinfo{pages}{1--13}.
\newblock
\showISBNx{978-1-4503-6708-0}
\href{https://doi.org/10.1145/3313831.3376785}{doi:\nolinkurl{10.1145/3313831.3376785}}


\bibitem[Lai et~al\mbox{.}(2021)]%
        {laiScienceHumanAIDecision2021}
\bibfield{author}{\bibinfo{person}{Vivian Lai}, \bibinfo{person}{Chacha Chen}, \bibinfo{person}{Q.~Vera Liao}, \bibinfo{person}{Alison {Smith-Renner}}, {and} \bibinfo{person}{Chenhao Tan}.} \bibinfo{year}{2021}\natexlab{}.
\newblock \bibinfo{title}{Towards a {{Science}} of {{Human-AI Decision Making}}: {{A Survey}} of {{Empirical Studies}}}.
\newblock \bibinfo{howpublished}{https://arxiv.org/abs/2112.11471v1}.
\newblock


\bibitem[Lee(2008)]%
        {leeSocialIdentityModel2008}
\bibfield{author}{\bibinfo{person}{Eun-Ju Lee}.} \bibinfo{year}{2008}\natexlab{}.
\newblock \showarticletitle{{Social Identity Model of Deindividuation Effects: Theoretical Implications and Future Directions}}.
\newblock \bibinfo{journal}{\emph{Communication Theories}} \bibinfo{volume}{4}, \bibinfo{number}{1} (\bibinfo{date}{June} \bibinfo{year}{2008}), \bibinfo{pages}{7--31}.
\newblock
\showISSN{1738-7221}


\bibitem[Liao and Sundar(2022)]%
        {liaoDesigningResponsibleTrust2022}
\bibfield{author}{\bibinfo{person}{Q.Vera Liao} {and} \bibinfo{person}{S.~Shyam Sundar}.} \bibinfo{year}{2022}\natexlab{}.
\newblock \showarticletitle{Designing for {{Responsible Trust}} in {{AI Systems}}: {{A Communication Perspective}}}. In \bibinfo{booktitle}{\emph{Proceedings of the 2022 {{ACM Conference}} on {{Fairness}}, {{Accountability}}, and {{Transparency}}}} \emph{(\bibinfo{series}{{{FAccT}} '22})}. \bibinfo{publisher}{Association for Computing Machinery}, \bibinfo{address}{New York, NY, USA}, \bibinfo{pages}{1257--1268}.
\newblock
\showISBNx{978-1-4503-9352-2}
\href{https://doi.org/10.1145/3531146.3533182}{doi:\nolinkurl{10.1145/3531146.3533182}}


\bibitem[Liu et~al\mbox{.}(2023)]%
        {liuLostMiddleHow2023}
\bibfield{author}{\bibinfo{person}{Nelson~F. Liu}, \bibinfo{person}{Kevin Lin}, \bibinfo{person}{John Hewitt}, \bibinfo{person}{Ashwin Paranjape}, \bibinfo{person}{Michele Bevilacqua}, \bibinfo{person}{Fabio Petroni}, {and} \bibinfo{person}{Percy Liang}.} \bibinfo{year}{2023}\natexlab{}.
\newblock \bibinfo{title}{Lost in the {{Middle}}: {{How Language Models Use Long Contexts}}}.
\newblock
\href{https://doi.org/10.48550/arXiv.2307.03172}{doi:\nolinkurl{10.48550/arXiv.2307.03172}}
\showeprint[arxiv]{2307.03172}~[cs]


\bibitem[Liu et~al\mbox{.}(2024)]%
        {liuProactiveConversationalAgents2024}
\bibfield{author}{\bibinfo{person}{Xingyu~Bruce Liu}, \bibinfo{person}{Shitao Fang}, \bibinfo{person}{Weiyan Shi}, \bibinfo{person}{Chien-Sheng Wu}, \bibinfo{person}{Takeo Igarashi}, {and} \bibinfo{person}{Xiang~`Anthony' Chen}.} \bibinfo{year}{2024}\natexlab{}.
\newblock \bibinfo{title}{Proactive {{Conversational Agents}} with {{Inner Thoughts}}}.
\newblock
\href{https://doi.org/10.48550/arXiv.2501.00383}{doi:\nolinkurl{10.48550/arXiv.2501.00383}}
\showeprint[arxiv]{2501.00383}~[cs]


\bibitem[Lu and Liao(2022)]%
        {luSurveyGroupDecision2022}
\bibfield{author}{\bibinfo{person}{Keyu Lu} {and} \bibinfo{person}{Huchang Liao}.} \bibinfo{year}{2022}\natexlab{}.
\newblock \showarticletitle{A Survey of Group Decision Making Methods in Healthcare Industry 4.0: Bibliometrics, Applications, and Directions}.
\newblock \bibinfo{journal}{\emph{Applied Intelligence (Dordrecht, Netherlands)}} \bibinfo{volume}{52}, \bibinfo{number}{12} (\bibinfo{year}{2022}), \bibinfo{pages}{13689--13713}.
\newblock
\showISSN{0924-669X}
\href{https://doi.org/10.1007/s10489-021-02909-y}{doi:\nolinkurl{10.1007/s10489-021-02909-y}}


\bibitem[Ma et~al\mbox{.}(2024)]%
        {maRecommenderExploratoryStudy2024}
\bibfield{author}{\bibinfo{person}{Shuai Ma}, \bibinfo{person}{Chenyi Zhang}, \bibinfo{person}{Xinru Wang}, \bibinfo{person}{Xiaojuan Ma}, {and} \bibinfo{person}{Ming Yin}.} \bibinfo{year}{2024}\natexlab{}.
\newblock \bibinfo{title}{Beyond {{Recommender}}: {{An Exploratory Study}} of the {{Effects}} of {{Different AI Roles}} in {{AI-Assisted Decision Making}}}.
\newblock
\href{https://doi.org/10.48550/arXiv.2403.01791}{doi:\nolinkurl{10.48550/arXiv.2403.01791}}
\showeprint[arxiv]{2403.01791}~[cs]


\bibitem[MacDougall and Baum(1997)]%
        {macdougallDevilsAdvocateStrategy1997}
\bibfield{author}{\bibinfo{person}{Colin MacDougall} {and} \bibinfo{person}{Frances Baum}.} \bibinfo{year}{1997}\natexlab{}.
\newblock \showarticletitle{The {{Devil}}'s {{Advocate}}: {{A Strategy}} to {{Avoid Groupthink}} and {{Stimulate Discussion}} in {{Focus Groups}}}.
\newblock \bibinfo{journal}{\emph{Qualitative Health Research}} \bibinfo{volume}{7}, \bibinfo{number}{4} (\bibinfo{date}{Nov.} \bibinfo{year}{1997}), \bibinfo{pages}{532--541}.
\newblock
\showISSN{1049-7323}
\href{https://doi.org/10.1177/104973239700700407}{doi:\nolinkurl{10.1177/104973239700700407}}


\bibitem[Maciejovsky et~al\mbox{.}(2013)]%
        {maciejovskyTeamsMakeYou2013}
\bibfield{author}{\bibinfo{person}{Boris Maciejovsky}, \bibinfo{person}{Matthias Sutter}, \bibinfo{person}{David~V. Budescu}, {and} \bibinfo{person}{Patrick Bernau}.} \bibinfo{year}{2013}\natexlab{}.
\newblock \showarticletitle{Teams {{Make You Smarter}}: {{How Exposure}} to {{Teams Improves Individual Decisions}} in {{Probability}} and {{Reasoning Tasks}}}.
\newblock \bibinfo{journal}{\emph{Management Science}} \bibinfo{volume}{59}, \bibinfo{number}{6} (\bibinfo{date}{June} \bibinfo{year}{2013}), \bibinfo{pages}{1255--1270}.
\newblock
\showISSN{0025-1909}
\href{https://doi.org/10.1287/mnsc.1120.1668}{doi:\nolinkurl{10.1287/mnsc.1120.1668}}


\bibitem[Mason(1969)]%
        {masonDialecticalApproachStrategic1969}
\bibfield{author}{\bibinfo{person}{Richard~O. Mason}.} \bibinfo{year}{1969}\natexlab{}.
\newblock \showarticletitle{A {{Dialectical Approach}} to {{Strategic Planning}}}.
\newblock \bibinfo{journal}{\emph{Management Science}} \bibinfo{volume}{15}, \bibinfo{number}{8} (\bibinfo{date}{April} \bibinfo{year}{1969}), \bibinfo{pages}{B--403}.
\newblock
\showISSN{0025-1909}
\href{https://doi.org/10.1287/mnsc.15.8.B403}{doi:\nolinkurl{10.1287/mnsc.15.8.B403}}


\bibitem[Mieczkowski et~al\mbox{.}(2021)]%
        {mieczkowskiAIMediatedCommunicationLanguage2021}
\bibfield{author}{\bibinfo{person}{Hannah Mieczkowski}, \bibinfo{person}{Jeffrey~T. Hancock}, \bibinfo{person}{Mor Naaman}, \bibinfo{person}{Malte Jung}, {and} \bibinfo{person}{Jess Hohenstein}.} \bibinfo{year}{2021}\natexlab{}.
\newblock \showarticletitle{{{AI-Mediated Communication}}: {{Language Use}} and {{Interpersonal Effects}} in a {{Referential Communication Task}}}.
\newblock \bibinfo{journal}{\emph{Proc. ACM Hum.-Comput. Interact.}} \bibinfo{volume}{5}, \bibinfo{number}{CSCW1} (\bibinfo{date}{April} \bibinfo{year}{2021}), \bibinfo{pages}{17:1--17:14}.
\newblock
\href{https://doi.org/10.1145/3449091}{doi:\nolinkurl{10.1145/3449091}}


\bibitem[Milana et~al\mbox{.}(2023)]%
        {milanaChatbotsAdvisersEffects2023}
\bibfield{author}{\bibinfo{person}{Federico Milana}, \bibinfo{person}{Enrico Costanza}, {and} \bibinfo{person}{Joel~E Fischer}.} \bibinfo{year}{2023}\natexlab{}.
\newblock \showarticletitle{Chatbots as {{Advisers}}: The {{Effects}} of {{Response Variability}} and {{Reply Suggestion Buttons}}}. In \bibinfo{booktitle}{\emph{Proceedings of the 5th {{International Conference}} on {{Conversational User Interfaces}}}} \emph{(\bibinfo{series}{{{CUI}} '23})}. \bibinfo{publisher}{Association for Computing Machinery}, \bibinfo{address}{New York, NY, USA}, \bibinfo{pages}{1--10}.
\newblock
\showISBNx{9798400700149}
\href{https://doi.org/10.1145/3571884.3597132}{doi:\nolinkurl{10.1145/3571884.3597132}}


\bibitem[Moscovici and Lage(1976)]%
        {moscoviciStudiesSocialInfluence1976}
\bibfield{author}{\bibinfo{person}{Serge Moscovici} {and} \bibinfo{person}{Elisabeth Lage}.} \bibinfo{year}{1976}\natexlab{}.
\newblock \showarticletitle{Studies in Social Influence {{III}}: {{Majority}} versus Minority Influence in a Group}.
\newblock \bibinfo{journal}{\emph{European Journal of Social Psychology}} \bibinfo{volume}{6}, \bibinfo{number}{2} (\bibinfo{year}{1976}), \bibinfo{pages}{149--174}.
\newblock
\showISSN{1099-0992}
\href{https://doi.org/10.1002/ejsp.2420060202}{doi:\nolinkurl{10.1002/ejsp.2420060202}}


\bibitem[Nass et~al\mbox{.}(1994)]%
        {nassComputersAreSocial1994}
\bibfield{author}{\bibinfo{person}{Clifford Nass}, \bibinfo{person}{Jonathan Steuer}, {and} \bibinfo{person}{Ellen~R. Tauber}.} \bibinfo{year}{1994}\natexlab{}.
\newblock \showarticletitle{Computers Are Social Actors}. In \bibinfo{booktitle}{\emph{Proceedings of the {{SIGCHI Conference}} on {{Human Factors}} in {{Computing Systems}}}} \emph{(\bibinfo{series}{{{CHI}} '94})}. \bibinfo{publisher}{Association for Computing Machinery}, \bibinfo{address}{New York, NY, USA}, \bibinfo{pages}{72--78}.
\newblock
\showISBNx{978-0-89791-650-9}
\href{https://doi.org/10.1145/191666.191703}{doi:\nolinkurl{10.1145/191666.191703}}


\bibitem[Natarajan et~al\mbox{.}(2024)]%
        {natarajanHumanintheloopAIintheloopAutomate2024}
\bibfield{author}{\bibinfo{person}{Sriraam Natarajan}, \bibinfo{person}{Saurabh Mathur}, \bibinfo{person}{Sahil Sidheekh}, \bibinfo{person}{Wolfgang Stammer}, {and} \bibinfo{person}{Kristian Kersting}.} \bibinfo{year}{2024}\natexlab{}.
\newblock \bibinfo{title}{Human-in-the-Loop or {{AI-in-the-loop}}? {{Automate}} or {{Collaborate}}?}
\newblock
\href{https://doi.org/10.48550/arXiv.2412.14232}{doi:\nolinkurl{10.48550/arXiv.2412.14232}}
\showeprint[arxiv]{2412.14232}~[cs]


\bibitem[Nemeth et~al\mbox{.}(2001)]%
        {nemethDevilsAdvocateAuthentic2001}
\bibfield{author}{\bibinfo{person}{Charlan Nemeth}, \bibinfo{person}{Keith Brown}, {and} \bibinfo{person}{John Rogers}.} \bibinfo{year}{2001}\natexlab{}.
\newblock \showarticletitle{Devil's Advocate versus Authentic Dissent: Stimulating Quantity and Quality}.
\newblock \bibinfo{journal}{\emph{European Journal of Social Psychology}} \bibinfo{volume}{31}, \bibinfo{number}{6} (\bibinfo{year}{2001}), \bibinfo{pages}{707--720}.
\newblock
\showISSN{1099-0992}
\href{https://doi.org/10.1002/ejsp.58}{doi:\nolinkurl{10.1002/ejsp.58}}


\bibitem[Poddar et~al\mbox{.}(2023)]%
        {poddarAIWritingAssistants2023}
\bibfield{author}{\bibinfo{person}{Ritika Poddar}, \bibinfo{person}{Rashmi Sinha}, \bibinfo{person}{Mor Naaman}, {and} \bibinfo{person}{Maurice Jakesch}.} \bibinfo{year}{2023}\natexlab{}.
\newblock \showarticletitle{{{AI Writing Assistants Influence Topic Choice}} in {{Self-Presentation}}}. In \bibinfo{booktitle}{\emph{Extended {{Abstracts}} of the 2023 {{CHI Conference}} on {{Human Factors}} in {{Computing Systems}}}} \emph{(\bibinfo{series}{{{CHI EA}} '23})}. \bibinfo{publisher}{Association for Computing Machinery}, \bibinfo{address}{New York, NY, USA}, \bibinfo{pages}{1--6}.
\newblock
\showISBNx{978-1-4503-9422-2}
\href{https://doi.org/10.1145/3544549.3585893}{doi:\nolinkurl{10.1145/3544549.3585893}}


\bibitem[Robertson et~al\mbox{.}(2021)]%
        {robertsonCantReplyThat2021}
\bibfield{author}{\bibinfo{person}{Ronald~E Robertson}, \bibinfo{person}{Alexandra Olteanu}, \bibinfo{person}{Fernando Diaz}, \bibinfo{person}{Milad Shokouhi}, {and} \bibinfo{person}{Peter Bailey}.} \bibinfo{year}{2021}\natexlab{}.
\newblock \showarticletitle{``{{I Can}}'t {{Reply}} with {{That}}'': {{Characterizing Problematic Email Reply Suggestions}}}. In \bibinfo{booktitle}{\emph{Proceedings of the 2021 {{CHI Conference}} on {{Human Factors}} in {{Computing Systems}}}} \emph{(\bibinfo{series}{{{CHI}} '21})}. \bibinfo{publisher}{Association for Computing Machinery}, \bibinfo{address}{New York, NY, USA}, \bibinfo{pages}{1--18}.
\newblock
\showISBNx{978-1-4503-8096-6}
\href{https://doi.org/10.1145/3411764.3445557}{doi:\nolinkurl{10.1145/3411764.3445557}}


\bibitem[{Schulz-Hardt} et~al\mbox{.}(2002)]%
        {schulz-hardtProductiveConflictGroup2002}
\bibfield{author}{\bibinfo{person}{Stefan {Schulz-Hardt}}, \bibinfo{person}{Marc Jochims}, {and} \bibinfo{person}{Dieter Frey}.} \bibinfo{year}{2002}\natexlab{}.
\newblock \showarticletitle{Productive Conflict in Group Decision Making: Genuine and Contrived Dissent as Strategies to Counteract Biased Information Seeking}.
\newblock \bibinfo{journal}{\emph{Organizational Behavior and Human Decision Processes}} \bibinfo{volume}{88}, \bibinfo{number}{2} (\bibinfo{date}{July} \bibinfo{year}{2002}), \bibinfo{pages}{563--586}.
\newblock
\showISSN{0749-5978}
\href{https://doi.org/10.1016/S0749-5978(02)00001-8}{doi:\nolinkurl{10.1016/S0749-5978(02)00001-8}}


\bibitem[Schweiger et~al\mbox{.}(1986)]%
        {schweigerGroupApproachesImproving1986}
\bibfield{author}{\bibinfo{person}{David~M. Schweiger}, \bibinfo{person}{William~R. Sandberg}, {and} \bibinfo{person}{James~W. Ragan}.} \bibinfo{year}{1986}\natexlab{}.
\newblock \showarticletitle{Group {{Approaches}} for {{Improving Strategic Decision Making}}: {{A Comparative Analysis}} of {{Dialectical Inquiry}}, {{Devil}}'s {{Advocacy}}, and {{Consensus}}}.
\newblock \bibinfo{journal}{\emph{Academy of Management Journal}} \bibinfo{volume}{29}, \bibinfo{number}{1} (\bibinfo{date}{March} \bibinfo{year}{1986}), \bibinfo{pages}{51--71}.
\newblock
\showISSN{0001-4273}
\href{https://doi.org/10.5465/255859}{doi:\nolinkurl{10.5465/255859}}


\bibitem[Schwenk and Valacich(1994)]%
        {schwenkEffectsDevilsAdvocacy1994}
\bibfield{author}{\bibinfo{person}{Charles Schwenk} {and} \bibinfo{person}{Joseph~S. Valacich}.} \bibinfo{year}{1994}\natexlab{}.
\newblock \showarticletitle{Effects of {{Devil}}{$\prime$}s {{Advocacy}} and {{Dialectical Inquiry}} on {{Individuals}} versus {{Groups}}}.
\newblock \bibinfo{journal}{\emph{Organizational Behavior and Human Decision Processes}} \bibinfo{volume}{59}, \bibinfo{number}{2} (\bibinfo{date}{Aug.} \bibinfo{year}{1994}), \bibinfo{pages}{210--222}.
\newblock
\showISSN{0749-5978}
\href{https://doi.org/10.1006/obhd.1994.1057}{doi:\nolinkurl{10.1006/obhd.1994.1057}}


\bibitem[Sebo et~al\mbox{.}(2020)]%
        {seboRobotsGroupsTeams2020}
\bibfield{author}{\bibinfo{person}{Sarah Sebo}, \bibinfo{person}{Brett Stoll}, \bibinfo{person}{Brian Scassellati}, {and} \bibinfo{person}{Malte~F. Jung}.} \bibinfo{year}{2020}\natexlab{}.
\newblock \showarticletitle{Robots in {{Groups}} and {{Teams}}: {{A Literature Review}}}.
\newblock \bibinfo{journal}{\emph{Proc. ACM Hum.-Comput. Interact.}} \bibinfo{volume}{4}, \bibinfo{number}{CSCW2} (\bibinfo{date}{Oct.} \bibinfo{year}{2020}), \bibinfo{pages}{176:1--176:36}.
\newblock
\href{https://doi.org/10.1145/3415247}{doi:\nolinkurl{10.1145/3415247}}


\bibitem[Sharma et~al\mbox{.}(2016)]%
        {sharmaGroupDecisionMaking2016}
\bibfield{author}{\bibinfo{person}{Vishakha Sharma}, \bibinfo{person}{Andrew Stranieri}, \bibinfo{person}{Frada Burstein}, \bibinfo{person}{Jim Warren}, \bibinfo{person}{Sharon Daly}, \bibinfo{person}{Louise Patterson}, \bibinfo{person}{John Yearwood}, {and} \bibinfo{person}{Alan Wolff}.} \bibinfo{year}{2016}\natexlab{}.
\newblock \showarticletitle{Group Decision Making in Health Care: {{A}} Case Study of Multidisciplinary Meetings}.
\newblock \bibinfo{journal}{\emph{Journal of Decision Systems}} \bibinfo{volume}{25}, \bibinfo{number}{sup1} (\bibinfo{date}{June} \bibinfo{year}{2016}), \bibinfo{pages}{476--485}.
\newblock
\showISSN{1246-0125}
\href{https://doi.org/10.1080/12460125.2016.1187388}{doi:\nolinkurl{10.1080/12460125.2016.1187388}}


\bibitem[Stasser et~al\mbox{.}(1989)]%
        {stasserInformationSamplingStructured1989}
\bibfield{author}{\bibinfo{person}{Garold Stasser}, \bibinfo{person}{Laurie~A. Taylor}, {and} \bibinfo{person}{Coleen Hanna}.} \bibinfo{year}{1989}\natexlab{}.
\newblock \showarticletitle{Information Sampling in Structured and Unstructured Discussions of Three- and Six-Person Groups}.
\newblock \bibinfo{journal}{\emph{Journal of Personality and Social Psychology}} \bibinfo{volume}{57}, \bibinfo{number}{1} (\bibinfo{year}{1989}), \bibinfo{pages}{67--78}.
\newblock
\showISSN{1939-1315}
\href{https://doi.org/10.1037/0022-3514.57.1.67}{doi:\nolinkurl{10.1037/0022-3514.57.1.67}}


\bibitem[Staufer et~al\mbox{.}(2024)]%
        {stauferSilencingRiskNot2024}
\bibfield{author}{\bibinfo{person}{Dimitri Staufer}, \bibinfo{person}{Frank Pallas}, {and} \bibinfo{person}{Bettina Berendt}.} \bibinfo{year}{2024}\natexlab{}.
\newblock \showarticletitle{Silencing the {{Risk}}, {{Not}} the {{Whistle}}: {{A Semi-automated Text Sanitization Tool}} for {{Mitigating}} the {{Risk}} of {{Whistleblower Re-Identification}}}. In \bibinfo{booktitle}{\emph{Proceedings of the 2024 {{ACM Conference}} on {{Fairness}}, {{Accountability}}, and {{Transparency}}}} \emph{(\bibinfo{series}{{{FAccT}} '24})}. \bibinfo{publisher}{Association for Computing Machinery}, \bibinfo{address}{New York, NY, USA}, \bibinfo{pages}{733--745}.
\newblock
\showISBNx{9798400704505}
\href{https://doi.org/10.1145/3630106.3658936}{doi:\nolinkurl{10.1145/3630106.3658936}}


\bibitem[Tanprasert et~al\mbox{.}(2024)]%
        {tanprasertDebateChatbotsFacilitate2024}
\bibfield{author}{\bibinfo{person}{Thitaree Tanprasert}, \bibinfo{person}{Sidney~S Fels}, \bibinfo{person}{Luanne Sinnamon}, {and} \bibinfo{person}{Dongwook Yoon}.} \bibinfo{year}{2024}\natexlab{}.
\newblock \showarticletitle{Debate {{Chatbots}} to {{Facilitate Critical Thinking}} on {{YouTube}}: {{Social Identity}} and {{Conversational Style Make A Difference}}}. In \bibinfo{booktitle}{\emph{Proceedings of the {{CHI Conference}} on {{Human Factors}} in {{Computing Systems}}}} \emph{(\bibinfo{series}{{{CHI}} '24})}. \bibinfo{publisher}{Association for Computing Machinery}, \bibinfo{address}{New York, NY, USA}, \bibinfo{pages}{1--24}.
\newblock
\showISBNx{9798400703300}
\href{https://doi.org/10.1145/3613904.3642513}{doi:\nolinkurl{10.1145/3613904.3642513}}


\bibitem[Tropman(2013)]%
        {tropmanEffectiveMeetingsImproving2013}
\bibfield{author}{\bibinfo{person}{John~E. Tropman}.} \bibinfo{year}{2013}\natexlab{}.
\newblock \bibinfo{booktitle}{\emph{Effective {{Meetings}}: {{Improving Group Decision Making}}}}.
\newblock \bibinfo{publisher}{SAGE Publications}.
\newblock
\showISBNx{978-1-4833-6564-0}


\bibitem[Uzzi et~al\mbox{.}(2013)]%
        {uzziAtypicalCombinationsScientific2013}
\bibfield{author}{\bibinfo{person}{Brian Uzzi}, \bibinfo{person}{Satyam Mukherjee}, \bibinfo{person}{Michael Stringer}, {and} \bibinfo{person}{Ben Jones}.} \bibinfo{year}{2013}\natexlab{}.
\newblock \showarticletitle{Atypical {{Combinations}} and {{Scientific Impact}}}.
\newblock \bibinfo{journal}{\emph{Science}} \bibinfo{volume}{342}, \bibinfo{number}{6157} (\bibinfo{date}{Oct.} \bibinfo{year}{2013}), \bibinfo{pages}{468--472}.
\newblock
\href{https://doi.org/10.1126/science.1240474}{doi:\nolinkurl{10.1126/science.1240474}}


\bibitem[Vogler and Robinson(2016)]%
        {voglerTeamBasedTestingImproves2016}
\bibfield{author}{\bibinfo{person}{Jane~S. Vogler} {and} \bibinfo{person}{Daniel~H. Robinson}.} \bibinfo{year}{2016}\natexlab{}.
\newblock \showarticletitle{Team-{{Based Testing Improves Individual Learning}}}.
\newblock \bibinfo{journal}{\emph{The Journal of Experimental Education}} \bibinfo{volume}{84}, \bibinfo{number}{4} (\bibinfo{date}{Oct.} \bibinfo{year}{2016}), \bibinfo{pages}{787--803}.
\newblock
\showISSN{0022-0973}
\href{https://doi.org/10.1080/00220973.2015.1134420}{doi:\nolinkurl{10.1080/00220973.2015.1134420}}


\bibitem[Wang et~al\mbox{.}(2022)]%
        {wangUnderstandingDesignSpace2022}
\bibfield{author}{\bibinfo{person}{Qiaosi Wang}, \bibinfo{person}{Ida Camacho}, \bibinfo{person}{Shan Jing}, {and} \bibinfo{person}{Ashok~K. Goel}.} \bibinfo{year}{2022}\natexlab{}.
\newblock \showarticletitle{Understanding the {{Design Space}} of {{AI-Mediated Social Interaction}} in {{Online Learning}}: {{Challenges}} and {{Opportunities}}}.
\newblock \bibinfo{journal}{\emph{Proc. ACM Hum.-Comput. Interact.}} \bibinfo{volume}{6}, \bibinfo{number}{CSCW1} (\bibinfo{date}{April} \bibinfo{year}{2022}), \bibinfo{pages}{130:1--130:26}.
\newblock
\href{https://doi.org/10.1145/3512977}{doi:\nolinkurl{10.1145/3512977}}


\bibitem[Wei et~al\mbox{.}(2023)]%
        {weiMultiPartyChatConversational2023}
\bibfield{author}{\bibinfo{person}{Jimmy Wei}, \bibinfo{person}{Kurt Shuster}, \bibinfo{person}{Arthur Szlam}, \bibinfo{person}{Jason Weston}, \bibinfo{person}{Jack Urbanek}, {and} \bibinfo{person}{Mojtaba Komeili}.} \bibinfo{year}{2023}\natexlab{}.
\newblock \bibinfo{title}{Multi-{{Party Chat}}: {{Conversational Agents}} in {{Group Settings}} with {{Humans}} and {{Models}}}.
\newblock
\href{https://doi.org/10.48550/arXiv.2304.13835}{doi:\nolinkurl{10.48550/arXiv.2304.13835}}
\showeprint[arxiv]{2304.13835}~[cs]


\bibitem[Wold(1986)]%
        {woldGroupDecisionMaking1986}
\bibfield{author}{\bibinfo{person}{J.~E. Wold}.} \bibinfo{year}{1986}\natexlab{}.
\newblock \showarticletitle{Group Decision Making: Teaching the Process--an Introductory {{Guided Design}} Project}.
\newblock \bibinfo{journal}{\emph{The Journal of Nursing Education}} \bibinfo{volume}{25}, \bibinfo{number}{9} (\bibinfo{date}{Nov.} \bibinfo{year}{1986}), \bibinfo{pages}{388--389}.
\newblock
\showISSN{0148-4834}
\href{https://doi.org/10.3928/0148-4834-19861101-10}{doi:\nolinkurl{10.3928/0148-4834-19861101-10}}


\bibitem[Xuetao et~al\mbox{.}(2009)]%
        {xuetaoImpactAgentsAnswers2009}
\bibfield{author}{\bibinfo{person}{Mao Xuetao}, \bibinfo{person}{Fran{\c c}ois Bouchet}, {and} \bibinfo{person}{Jean-Paul Sansonnet}.} \bibinfo{year}{2009}\natexlab{}.
\newblock \showarticletitle{Impact of Agent's Answers Variability on Its Believability and Human-Likeness and Consequent Chatbot Improvements}. In \bibinfo{booktitle}{\emph{Proceedings of {{AISB}}}}.
\newblock


\bibitem[Zhang et~al\mbox{.}(2025)]%
        {zhangBreakingBarriersBuilding2025}
\bibfield{author}{\bibinfo{person}{Zihan Zhang}, \bibinfo{person}{Black Sun}, {and} \bibinfo{person}{Pengcheng An}.} \bibinfo{year}{2025}\natexlab{}.
\newblock \bibinfo{title}{Breaking {{Barriers}} or {{Building Dependency}}? {{Exploring Team-LLM Collaboration}} in {{AI-infused Classroom Debate}}}.
\newblock
\href{https://doi.org/10.48550/arXiv.2501.09165}{doi:\nolinkurl{10.48550/arXiv.2501.09165}}
\showeprint[arxiv]{2501.09165}~[cs]


\bibitem[Zheng et~al\mbox{.}(2023)]%
        {zhengCompetentRigidIdentifying2023}
\bibfield{author}{\bibinfo{person}{Chengbo Zheng}, \bibinfo{person}{Yuheng Wu}, \bibinfo{person}{Chuhan Shi}, \bibinfo{person}{Shuai Ma}, \bibinfo{person}{Jiehui Luo}, {and} \bibinfo{person}{Xiaojuan Ma}.} \bibinfo{year}{2023}\natexlab{}.
\newblock \showarticletitle{Competent but {{Rigid}}: {{Identifying}} the {{Gap}} in {{Empowering AI}} to {{Participate Equally}} in {{Group Decision-Making}}}. In \bibinfo{booktitle}{\emph{Proceedings of the 2023 {{CHI Conference}} on {{Human Factors}} in {{Computing Systems}}}} \emph{(\bibinfo{series}{{{CHI}} '23})}. \bibinfo{publisher}{Association for Computing Machinery}, \bibinfo{address}{New York, NY, USA}, \bibinfo{pages}{1--19}.
\newblock
\showISBNx{978-1-4503-9421-5}
\href{https://doi.org/10.1145/3544548.3581131}{doi:\nolinkurl{10.1145/3544548.3581131}}


\end{thebibliography}

\end{document}